\documentclass[%
 preprint,
nobibnotes,
 amsmath,amssymb,
 aps,prl,
]{revtex4-1}

\usepackage{graphicx,epstopdf,amssymb,amsfonts,amsmath,amsthm,array,
mathrsfs,amscd,mathtools}
\usepackage{dcolumn}
\usepackage{bm}
\usepackage{hyperref}

\def\be{\begin{equation}}\def\ee{\end{equation}}
\def\cvp{\raise 2pt\hbox{,}} 
 \def\tr{\mathop{\rm tr}\nolimits}

 \def\d{{\rm d}}\def\nn{{\cal
N}}

\def\gs{g_{\text s}}\def\ls{\ell_{\text s}}

\def\1{\mathbb{I}}

\def\iin{i_{1}\cdots i_{n}}

\def\imath#1#2#3{{\it Invent math }{\bf #1} (#2) #3}

\begin{document}


\title{Emergent D4-Brane Background from D-Particles
}

\author{Frank Ferrari}
 \email{frank.ferrari@ulb.ac.be}
 \author{Micha Moskovic}
 \email{micha.moskovic@ulb.ac.be}
 \affiliation{Service de 
 Physique Th\'eorique et
 Math\'ematique, Universit\'e Libre de Bruxelles and International
 Solvay Institutes, Campus de la Plaine, CP 231, B-1050 Bruxelles,
 Belgium}

\date{\today}

\begin{abstract}

We show that the solution of a pre-geometric strongly coupled quantum mechanical model describing $K$ D-particles in the presence of $N$ D4-branes in type IIA string theory, at fixed $K$ and large $N$, yields an effective action describing the motion of the $K$ D-particles in a classical ten-dimensional curved spacetime. By comparing the effective action with the non-abelian D-brane action in an arbitrary supergravity background, 
we identify the metric, dilaton and Ramond-Ramond fields and find a precise match with the near-horizon D4-brane geometry.

\end{abstract}

\maketitle


%
\section{Introduction}
\label{IntroSec}

A modern paradigm in quantum gravity, beautifully realized in the AdS/CFT correspondence \cite{Maldacena:1997re,*Gubser:1998bc,*Witten:1998qj,*Aharony:1999ti}, is to view space and gravity as emerging from an underlying, pre-geometric, strongly coupled quantum mechanical model. In the standard set-up, gauge theory correlators are related to the scattering of closed string modes off a large number $N$ of background branes. The closed string modes move in a dual curved ten-dimensional spacetime. This geometry is emergent from the point of view of the gauge theory, which lives on the background brane worldvolume. This framework has been extraordinarily fruitful to study strongly coupled gauge theories using gravity. Unfortunately, learning about gravity from large $N$ gauge theory has proven to be much more difficult. It is highly non-trivial to see the emerging geometry from microscopic gauge theory calculations.

Recently, it was proposed in \cite{Ferrari:2012nw} to use the scattering of probe D-branes off background branes to build calculable models of emergent space. A crucial point emphasized in \cite{Ferrari:2012nw} is that the pre-geometric models describing such scatterings are akin to vector models. A sum over an infinite set of planar diagrams, with any number of loops, contributing to the leading large $N$ approximation, can then always be performed. In favorable cases with conformal invariance and/or supersymmetry, this infinite sum of diagrams can yield exact terms in the effective action. Quite remarkably, this action describes, 
in principle, the motion of the probe D-branes in the emerging classical ten-dimensional geometry sourced by the background branes. 
By comparing with the non-abelian D-brane action in arbitrary backgrounds \cite{Myers:1999ps,*Taylor:1999gq,*Taylor:1999pr}, one can then derive the supergravity solution. These ideas have been applied successfully to the case of D-instantons in a variety of examples \cite{Ferrari:2012nw,Ferrari:2013pq}. 

The purpose of the present Letter is to carry out this program a significant step forward, by considering quantum mechanical particles as probes. We start from a pre-geometric quantum mechanics, in which only four space dimensions $X^{\mu}$, $1\leq\mu\leq 4$, are present and which contains additional ``abstract'' vector-like degrees of freedom $(q_{f},\tilde q^{f})$, $1\leq f\leq N$, together with fermionic superpartners. This model is motivated by the theory of D-particles moving on D4-branes in type IIA string theory. The $X^{\mu}$ correspond to the worldvolume space coordinates and the $(q,\tilde q)$ describe the interactions between the D-particles and the D4-branes. At large $N$, we show that the solution of the model is described by a classical action $S_{\text{eff}}(X^{\mu},Y^{A})$. The fluctuations of both the $X^{\mu}$ and the new variables $Y^{A}$, $1\leq A\leq 5$, which are composite in terms of the pre-geometric degrees of freedom $(q,\tilde q)$, are suppressed at large $N$. This demonstrates that the original large $N$ strongly coupled pre-geometric quantum mechanics is equivalent to the classical mechanics of 
D-particles moving in an emergent ten-dimensional curved background with space coordinates $(X^{\mu},Y^{A})$. 
By analyzing in detail some of the terms in the action $S_{\text{eff}}$, we can actually read off explicitly the non-vanishing ten-dimensional background fields. 
The metric, dilaton and Ramond-Ramond forms are found to match precisely the near-horizon D4-brane solution of type IIA supergravity \cite{Horowitz:1991cd,*Itzhaki:1998dd}.

\section{\label{lagsec} The quantum mechanical model}

The quantum mechanical system that we consider models the dynamics of $K$ D-particles interacting with $N$ coinciding D4-branes in type IIA string theory. It is related by T-duality to the D(-1)/D3 system considered in \cite{Ferrari:2012nw} and can be interpreted as a $\text{U}(K)$ gauged  supersymmetric quantum mechanics on the ADHM instanton moduli space. It can be obtained as in \cite{Aharony:1997th,*Berkooz:1999iz,Ferrari:2012nw} from a scaling limit, associated with the near-horizon limit in the background of the D4-branes, of the dimensionally reduced $\text{U}(K)$ $\nn=1$ super Yang-Mills theory with one adjoint and $N$ fundamental hypermultiplets from six to one dimension.

For convenience, we work in Euclidean signature. The model preserves eight supercharges. It has a $\text{U}(K)$ gauge symmetry, a $\text{U}(N)$ flavor symmetry and a $\text{SU}(2)_{+}\times\text{SU}(2)_{-}\times\text{Spin}(5)$ global symmetry corresponding to $\text{SO}(4)\times\text{SO(5)}$ rotations transverse to the worldlines of the D-particles and preserving the configuration of the background D4-branes.
The fundamental degrees of freedom in the adjoint of $\text{U}(K)$, associated with D0/D0 strings, are the D4-brane worldvolume matrix space coordinates $X_{\mu}$ and a $\text{SU}(2)_{-}$ doublet of $\text{Spin}(5)$ spinor superpartners $\psi_{\dot\alpha}$.
The D0/D4 strings yield additional degrees of freedom $(q_{\alpha},\chi)$ and $(\tilde q^{\alpha},\tilde\chi)$ in the fundamental and anti-fundamental representations of $\text{U}(N)\times\text{U}(K)$. The bosonic $q_{\alpha}$ and $\tilde q^{\alpha}$ are doublets of $\text{SU}(2)_{+}$ and the fermionic $\chi$ and $\tilde\chi$ are $\text{Spin}(5)$ spinors.
In terms of these variables, the action contains complicated four-fermion terms and the ADHM constraints must be imposed in the path integral.

The four-fermion terms can be greatly simplified by introducing a non-dynamical auxiliary $\text{SO}(5)$ vector $Y_{A}$ in the adjoint of $\text{U}(K)$, whereas the ADHM constraints can be implemented by using adjoint Lagrange multipliers $(D_{\mu\nu},\Lambda^{\alpha})$, where
the bosonic $D_{\mu\nu}$ is self-dual and the fermionic $\Lambda^{\alpha}$ is a $\text{SU}(2)_{+}$ doublet of $\text{Spin}(5)$ spinors.
We shall see that these auxiliary variables play a crucial r\^ole, both at the technical level to solve the model \cite{Coleman:1980nk,*ZinnJustin:1998cp,*Ferrari:2000wq,*Ferrari:2001jt,*Ferrari:2002gy} and for the physical interpretation of the solution in terms of an emerging geometry, in a way akin to the case of D-instantons \cite{Ferrari:2012nw,hep-th/9810243,*hep-th/9901128,*Akhmedov:1998pf}.

We are going to focus on the bosonic part of the effective action for the D-particles. We can thus set the adjoint fermions $\psi_{\dot\alpha}$ and $\Lambda^{\alpha}$ to zero. In terms of the $\text{SO}(5)$ Dirac matrices $\Gamma_{A}$, charge conjugation matrix $C$ and $\mathfrak{su}(2)_{+}$ generators $\sigma_{\mu\nu}$,
the microscopic Lagrangian we start from then reads
\begin{multline}
 L=\frac12\nabla\tilde  q^{\alpha}\nabla q_{\alpha}
  +\frac{i}2\tilde\chi C\nabla\chi
 +\frac{i}2\tilde q^{\alpha}D_{\mu\nu}\sigma_{\mu\nu\alpha}^{\phantom{\mu\nu\alpha}\beta}q_{\beta} 
  -\frac{i}{2\ls^2}\tilde\chi C\Gamma_{A}
  Y_{A}\chi
  +\frac1{2\ls^4}\tilde q^{\alpha}Y_AY_Aq_{\alpha} \\
  +\frac{\sqrt{2\pi}}{\gs\ls}\tr\Bigl(\! 1+\frac{1}{2} \nabla X_\mu \nabla X_\mu -\frac1{2\ls^4}[Y_A,X_\mu][Y_A,X_\mu] 
  +i[X_\mu,X_\nu]D_{\mu\nu} \Bigr)  
   \, ,
  \label{Ldef}
\end{multline}
where $\nabla$ is the worldline covariant derivative. The non-trivial normalization of the trace term in \eqref{Ldef} is fixed by the D-particle mass in type IIA string theory in terms of the string coupling $\gs$ and the string length $\ls$. 

Let us note that the full open-string description of the D0/D4 system includes a priori additional terms coupling the D4-brane worldvolume fields to the D-particle worldline variables. Similar terms for the D(-1)/D3 system have been discussed in \cite{Green:2000ke,*Billo:2002hm}. A comprehensive discussion of the effect of these terms, in a non-supersymmetric context, will be presented in \cite{frankonematrix}. Presently, with eight supersymmetries, these terms are expected to be irrelevant for some of the contributions in the effective action, on which we focus \cite{Ferrari:2012nw,Ferrari:2013pq}. In particular, for the kinetic term (see \eqref{gsolu}), this is consistent with the non-renormalization theorem discussed in \cite{Diaconescu:1997ut}.

The Lagrangian \eqref{Ldef} is pre-geometric, in the sense that there is no dynamical variable associated with the motion of the D-particles in directions transverse to the D4-branes. The interactions between the D-particles and the D4s are described ``abstractly'' by the pre-geometric variables $(q,\chi,\tilde q,\tilde\chi)$. Our goal is to show that the strong quantum effects generated by these non-geometric interactions literally create five new dimensions of space in which the D-particles can move. Moreover, we are going to prove that the resulting ten-dimensional spacetime behaves classically at large $N$, is curved and supports a non-trivial dilaton and three-form field, precisely matching the near-horizon D4-brane type IIA supergravity background \cite{Horowitz:1991cd,*Itzhaki:1998dd}. 

\section{\label{solsec} The solution of the model}

The model \eqref{Ldef} can be solved at large $N$ because the interacting degrees of freedom $(q,\chi,\tilde q,\tilde\chi)$ carry only one $\text{U}(N)$ index and thus are vector-like variables. The leading large $N$ Feynman diagrams are then multi-loop bubble diagrams which can always be summed up exactly. The well-known technical trick to elegantly perform this sum \cite{Coleman:1980nk,*ZinnJustin:1998cp,*Ferrari:2000wq,*Ferrari:2001jt,*Ferrari:2002gy} is to rewrite the complicated interactions between the vector degrees of freedom by introducing auxiliary fields, in such a way that the vector variables only appear quadratically in the action. This is exactly what we have done when writing \eqref{Ldef} in terms of $Y_{A}$ and $D_{\mu\nu}$. One then integrates exactly over these variables to obtain a non-local effective action $S_{\text{eff}}$ for the auxiliary fields. This effective action is automatically proportional to $N$. It can thus be treated classically when $N$ is large. The tree diagrams of $S_{\text{eff}}$ reproduce the leading large $N$ bubble diagrams of the original action. 

In our case, fixing the worldline $\text{U}(K)$ gauge invariance such that $\nabla=\d_{t}$ is the ordinary time derivative, the effective action reads
\be \label{Seff1} S_{\text{eff}}(X,Y,D)=\int\!\d t\, L_{\text{{tr}}} +N\bigl(\ln \Delta_B-\ln \Delta_F\bigr)\, ,\ee
where $L_{\text{tr}}$ is the trace term in \eqref{Ldef} and
\begin{align}\label{bosdet}
  \Delta_B&=\det\bigl( -\d_{t}^2+\ls^{-4}Y_AY_A+iD_{\mu\nu}\otimes\sigma_{\mu\nu} \bigr) \, , \\ \label{ferdet}
  \Delta_F&= \det\bigl(-i\d_{t}+i\ls^{-2}Y_A\otimes\Gamma_A\bigr)
\end{align}
are bosonic and fermionic functional determinants obtained by integrating out $(q,\tilde q)$ and $(\chi,\tilde\chi)$ respectively. We now claim that the classical action \eqref{Seff1} describes the motion of the D-particles in a ten-dimensional spacetime with coordinates $X_{\mu}$ and $Y_{A}$. In other words, the auxiliary variables $Y_{A}$, which have acquired dynamics through the quantum loops of the vector-like variables, can be interpreted as the coordinates of the emerging five-dimensional space transverse to the D4-branes.

To prove that this interpretation is sensible, we first integrate out $D_{\mu\nu}$ which, at large $N$, can be done by solving the saddle point equation
\be\label{Dsaddle} \delta S_{\text{eff}}/\delta D_{\mu\nu}(t) = 0\, .\ee
This yields a new effective action
\be\label{S10d} \tilde S_{\text{eff}}(X,Y) = S_{\text{eff}}\bigl(X,Y,\langle D\rangle\bigr)\ee
which will be compared in the next Section 
to the non-abelian action for D-particles in a general type IIA supergravity background. 

The action $\tilde S_{\text{eff}}$ can be most conveniently analyzed by expanding around time-independent diagonal configurations,
\begin{equation}
  X_\mu(t)=x_\mu\1_{K}+\ls^2\epsilon_\mu(t) \, , \quad Y_A(t)=y_A\1_{K}+\ls^2\epsilon_A(t) \, .
  \label{matrixexpansion}
\end{equation}
In this expansion, the determinants \eqref{bosdet} and \eqref{ferdet} can be computed by using the identity 
\begin{equation}
  \ln\det(M+\delta M)=\ln \det M-\sum_{k=1}^\infty \frac{(-1)^{k }}{k}\tr(M^{-1}\delta M)^k \, ,
  \label{Detexp}
\end{equation}
with 
\begin{equation}
  M=-\d_{t}^2+\ls^{-4}r^2  , \ \delta M=(2\ls^{-2}\vec y\cdot\vec\epsilon+\vec{\epsilon\,}^2)\1_2+i\left<D_{\mu\nu}\right>\sigma_{\mu\nu}
  \label{MB}
\end{equation}
in the bosonic case and
\begin{equation}
  M=-i\d_{t}+i\ls^{-2}y_A\Gamma_A \, , \quad \delta M=i\epsilon_A\Gamma_A 
  \label{MF}
\end{equation}
in the fermionic case. We use the notation
\be\label{radialdef} r^{2} = y_{A}y_{A}={\vec y\,}^{2}\ee
and $\vec y\cdot\vec\epsilon = y_{A}\epsilon_{A}$, etc. 
The corresponding bosonic and fermionic Green's functions read
\begin{align}
  G_B(t,t')& =\int \frac{\d\omega}{2\pi}\frac{e^{i\omega(t-t')}}{\omega^2+\ls^{-4}r^2} \, \cvp
  \label{defGB}\\
  G_F(t,t') &=\int \frac{\d\omega}{2\pi}\frac{e^{i\omega(t-t')}}{\omega^2+\ls^{-4}r^2}\bigl(\omega\1_4-i\ls^{-2}y_A\Gamma_A\bigr) \, .
  \label{GBdef}
\end{align}
The trace in \eqref{Detexp} involves both integrals over frequencies and traces over $\text{Spin}(5)$, $\text{SU}(2)_{+}$ and $\text{U}(K)$ indices.
At each order in the $\epsilon$ expansion, we can further expand in powers of the frequencies or, equivalently, in time derivatives.
The saddle-point equation \eqref{Dsaddle} can be solved similarly, both in the $\epsilon$ expansion \eqref{matrixexpansion} and in the derivative expansion.

If we write
\begin{equation}
  \tilde S_{\text{eff}}=\int\!\d t\,\sum_{p\ge0}L_{p} \, ,
  \label{Seffexp}
\end{equation}
where $L_p$ is of order $p$ in $\epsilon$, we find
\be\label{L0L1sol} L_{0}= K\frac{\sqrt{2\pi}}{\gs\ls}\, ,\quad L_{1} = 0\, ,\ee
which are simple consequences of supersymmetry. We have also computed $L_{2}$ and $L_{3}$ up to fourth order in derivatives,
\begin{align}\label{L2sol} L_{2} &=  \tr\Bigl( \frac{\sqrt{2\pi}\ls^3}{2g_s}\dot\epsilon_\mu\dot\epsilon_\mu+\frac{N\ls^6}{4r^3}\dot\epsilon_A\dot\epsilon_A \Bigr) + O\bigl(\ddot\epsilon^{2}\bigr)  \, , \\\label{L3sol}
  L_{3}&=-\frac{3N\ls^4}{4r^5}  \tr\bigl( \vec y\cdot\vec\epsilon\, \dot{\vec \epsilon}^{\,2} \bigr) + O\bigl(\epsilon\ddot\epsilon^2,\dot\epsilon^{2}\ddot\epsilon\bigr) \, , 
\end{align}
and $L_{4}$ and $L_{5}$ up to second order in derivatives,
\begin{align}\notag
 L_{4}&=-\tr \Bigl( \frac{\pi r^3}{g_s^2 N}[\epsilon_\mu,\epsilon_\nu][\epsilon_\mu,\epsilon_\nu]  +\frac{\ls^6N}{8r^3}[\epsilon_A,\epsilon_B][\epsilon_A,\epsilon_B]\\
 &\hskip 5.5cm+\frac{\sqrt{2\pi}\ls^3}{2g_s}[\epsilon_A,\epsilon_\mu][\epsilon_A,\epsilon_\mu] \Bigr) +O\bigl(\epsilon^{2}\dot\epsilon^2\bigr) 
 \, ,\label{L4} \\
  L_{5}&=-\frac{6\pi\ls^2r}{g_s^2N} \epsilon_{\mu\nu\rho\kappa}\tr\epsilon_\mu\epsilon_\nu\epsilon_\rho\epsilon_\kappa\vec y\cdot\vec\epsilon +\cdots+O\bigl(\epsilon^{3}\dot\epsilon^2\bigr) \mathrlap{.}
  \label{L5}
\end{align}
The $\cdots$ in \eqref{L5} are contributions to the action that are fixed in terms of \eqref{L4} by general consistency conditions \cite{Ferrari:2013pi}. We are now going to show that the terms \eqref{L0L1sol}--\eqref{L5} perfectly match with the expected form of the D-particle Lagrangian in a non-trivial background.

\section{\label{emergesec} The emergent geometry}

The non-abelian action for D-particles in an arbitrary background can be computed using formulas in \cite{Myers:1999ps,*Taylor:1999gq,*Taylor:1999pr}, see \cite{Ferrari:2013pi} for details. If we denote the space matrix coordinates by $Z^{i}$, $1\leq i\leq 9$, and expand $Z^{i} = z^{i} +\ls^{2} \epsilon^{i}$, then the Lagrangian, computed in the static gauge
\be\label{staticgauge} x^{d}=x^{10} = t\, ,\ee
can be conveniently written as a sum of terms with a fixed number of derivatives, 
\begin{multline}
\label{order0exp} L = \sum_{n\geq 0}\frac{1}{n!}\ls^{2n}
c_{\iin}^{(0)}(z,t)
\tr\epsilon^{i_{1}}\cdots\epsilon^{i_{n}}  +
\sum_{n\geq 0}\frac{1}{n!}\ls^{2(n+1)}
c^{(1)}_{\iin; k} (z,t) \tr\epsilon^{i_{1}}\cdots\epsilon^{i_{n}} \dot\epsilon^{k}\\ +
\ls^{4} c^{(2)}_{kl}(z,t)\tr\dot\epsilon^{k}\dot\epsilon^{l}+ \ls^{6}c^{(2)}_{i;kl}(z,t)\tr\epsilon^{i}\dot\epsilon^{k}\dot\epsilon^{l} + \cdots
\end{multline}
The $\cdots$ represent terms of higher orders in $\epsilon$ with two derivatives or with at least three derivatives.
The coefficients in \eqref{order0exp} can be expressed in terms of the type IIA supergravity fields. Our goal is to find a match between \eqref{order0exp} and the corresponding terms in our microscopically computed Lagrangian \eqref{L0L1sol}--\eqref{L5}, with $\epsilon^{i} \equiv (\epsilon_{\mu},\epsilon_{A})$. 
We are seeking a static $\text{SO}(4)\times\text{SO}(5)$ preserving background which has vanishing Neveu-Schwarz $B$ field and Ramond-Ramond one-form.
The coefficients in \eqref{order0exp} can then be naturally expressed in terms of
the following combinations of $G_{MN}$, $1\leq M,N\leq 10$, and dilaton $\phi$,
\be\label{gdef} g_{MN}  = e^{-2\phi}G_{MN}\, ,\quad 
\mathscr H_{ij} = \sqrt{g_{dd}}\biggl( \frac{g_{ij}}{g_{dd}} -
\frac{g_{di} g_{dj}}{g_{dd}^{2}}\biggr) \, .
\ee
If $C$ and $\tilde C$ are the Ramond-Ramond three-form potential and its dual five-form respectively, $\d\tilde C=i*\d C$ in the Euclidean, the explicit formulas we need read \cite{Ferrari:2013pi}
\begin{align}
\label{c0} c^{(0)}  &= \frac{\sqrt{2\pi}}{\ls} \sqrt{g_{dd}}\, ,\quad
c^{(1)}_{i;k}  = \frac{\sqrt{2\pi}}{\ls}\partial_{i}\bigl(g_{dk}/\sqrt{g_{dd}}\bigr)\, ,
\\  
\label{pot2D0a} c^{(2)}_{kl}  &= \frac{\sqrt{2\pi}}{2\ls} \mathscr H_{kl} \, , \\
\label{potD0b} c^{(0)}_{[ijk]}  &= \frac{3\sqrt{2\pi}}{2\ls^{3}}
\partial_{[i} C_{jk]d}\, ,  \\
c^{(0)}_{[ij][kl]}  &= -\frac{9\sqrt{2\pi}}{\ls^{5}}
g_{dd}^{3/2}e^{4\phi} \bigl(\mathscr H_{ik}\mathscr H_{jl} - \mathscr H_{jk}\mathscr H_{il}\bigr) \, ,\label{potD0d} \\
c^{(0)}_{[ijklm]}  &= -\frac{60i\sqrt{2\pi}}{\ls^{5}}\partial_{[i}
 \tilde C_{jklm]d} \, . \label{potD0e}
\end{align}
Other combinations of coefficients either vanish, consistently with the vanishing of some of the background fields, or are expressed in terms of \eqref{c0}--\eqref{potD0e} by solving the general consistency conditions discussed in \cite{Ferrari:2013pi}.

A single D-particle, $K=1$, probes only the metric $g_{MN}$. In this case, 
by comparing \eqref{c0} and \eqref{pot2D0a} with \eqref{L0L1sol} and \eqref{L2sol} and by noting that non-vanishing constant components $g_{dk}$ would be inconsistent with $\text{SO(4)}\times\text{SO}(5)$, we find
\be\label{gsolu} g_{MN}\, \d x^{M}\d x^{N} = \frac{1}{\gs^2}\Big(\d t^{2} + \d x_{\mu}\d x_{\mu} +
\frac{L^{3}}{r^{3}} {\d\vec y}^{\,2}\Big) \, ,\ee
with
\be\label{Length} L^{3} = \frac{N\gs\ls^{3}}{2\sqrt{2\pi}}\,\cdotp\ee
%
Using the full non-abelian action, $K>1$, we can get more information on the background. Indeed, the D-particles then couple to the Ramond-Ramond three-form through commutator terms. By comparing \eqref{potD0b} with \eqref{L3sol} and \eqref{potD0e} with \eqref{L5}, we can find $F_{4}=\d C$ or its dual $F_{6}$ unambiguously. Moreover, one can check that the double commutator term in the fourth order potential \eqref{L4} has precisely the correct structure to match with \eqref{potD0d}. Since $\mathscr H_{ij}$ is already known from the kinetic term \eqref{L2sol}, we can derive the dilaton profile from this term and then extract the string frame metric $\d s^{2}$ from 
\eqref{gsolu} and the first equation in \eqref{gdef}.
Overall, we get
\begin{align}
 & \d s^{2}=\frac{r^{3/2}}{L^{3/2}}(\d t^2+\d x_\mu\d x_\mu)+\frac{L^{3/2}}{r^{3/2}}\d{\vec y\,}^2 \, , \label{metric} \\
 & e^{\phi}=g_s \frac{r^{3/4}}{L^{3/4}} \, \cvp \label{dilaton} \\
  &  F_{4} = \frac{L^{3}}{8\gs r^{5}}\epsilon_{ABCDE}\,y_{E}\,\d y_{A}\wedge\cdots\wedge\d y_{D}\, .
\end{align}
This background is in perfect agreement with the near-horizon D4-brane background \cite{Horowitz:1991cd,*Itzhaki:1998dd}, including the relation between the supergravity length scale $L$ and string-theory parameters $g_s$ and $\ls$ and the correct normalization of the Ramond-Ramond form, consistently with the D4-brane charge in type IIA.

\section{Conclusion}
\label{ConcSec}

We have explicitly constructed a pre-geometric model for particles interacting with a large number of other degrees of freedom. The emergence of space and gravity are made possible by strong quantum mechanical effects in the pre-geometric description. The model we have solved is quite peculiar, in particular it preserves eight supersymmetries, but the ideas involved are very general and can be applied to a large number of examples, see e.g.\ \cite{emergentD1}. A comprehensive discussion of some of the many possible generalizations will appear in a related paper \cite{frankonematrix}. 

Being able to build solvable models of emergent space opens up very exciting possibilities. Outstanding problems for the future will be to build calculable pre-geometric descriptions of classically singular backgrounds, including black holes. This could bring interesting new perspectives on the controversial physics associated with such spacetimes.

We believe that the emergent space paradigm for quantum gravity will become increasingly convincing when more detailed solutions of explicit examples  become available. This point of view completely eliminates the old difficulties which superficially make quantum mechanics and gravity hard to reconcile. On the contrary, we can find space and gravity only as a consequence of quantum mechanics.

\begin{acknowledgments}

We would like to thank Antonin Rovai for collaboration at an early stage of this project.
This work is supported in part by the Belgian Fonds de la Recherche
Fondamentale Collective (grant 2.4655.07) and the Belgian Institut
Interuniversitaire des Sciences Nucl\'eaires (grants 4.4511.06 and 4.4514.08).
M.M.\ is a Research Fellow of the Belgian Fonds de la Recherche Scientifique - FNRS.

\end{acknowledgments}
\bibliographystyle{utphys}
\bibliography{emergentbiblio}
\end{document}